\documentclass[pdflatex,sn-mathphys]{sn-jnl}

\usepackage{xcolor}

\usepackage[utf8]{inputenc}
\usepackage{subfig}
\usepackage{makecell, cellspace, caption}

\usepackage{graphicx}
\usepackage{amsmath}
\usepackage{mathtools}

\usepackage{booktabs}

\usepackage{algorithm,algorithmicx,algpseudocode}
\usepackage{multirow}
\usepackage{multicol}
\usepackage[numbers]{natbib}

\usepackage{perpage} %the perpage package
\MakePerPage{footnote} %the perpage package

\newcommand{\STAB}[1]{\begin{tabular}{@{}c@{}}#1\end{tabular}}

\newcolumntype{C}[1]{>{\centering\arraybackslash}p{#1}}
\newcolumntype{P}[1]{>{\raggedright\arraybackslash}p{#1}}

\begin{document}

% suggested titles
% BERT-based Conflict Detection in SRS documents
% 
% \title{Adaptive Fine-tuning for Multiclass Requirement Classification}
\title[Multiclass Classification over Software Requirement Data]{Adaptive Fine-tuning for Multiclass Classification over Software Requirement Data}

\author[1]{\fnm{Savas} \sur{Yildirim}}
\author*[1]{\fnm{Mucahit} \sur{Cevik}}\email{mcevik@ryerson.ca}
\author[2]{\fnm{Devang} \sur{Parikh}} 
\author[1]{\fnm{Ayse} \sur{Basar}}

% \authormark{Yildirim et al.}

\affil[1]{
% \orgdiv{Data Science Lab}, 
\orgname{Toronto Metropolitan University}, \orgaddress{\street{44 Gerrard St E}, \city{Toronto}, \postcode{M5B 1G3}, \state{Ontario}, \country{Canada}}}

\affil[2]{
\orgname{IBM}, \orgaddress{\city{Cary}, \postcode{27709}, \state{North Carolina}, \country{USA}}}

% \corres{*Mucahit Cevik, Data Science Lab, Mechanical Industrial Engineering Department, Ryerson University, Toronto, Ontario, M5B 2K3, Canada. \email{mcevik@ryerson.ca}}

%%%%%%%%%%%%%%%%%%%%%%%%%%%%%%%%%%%%%%%%%%%%%%%%%%%%%%%%%%%%%%%%%%%%%%%%%%%%%%%%%
% \abstract[Abstract]{%context
\abstract{
The analysis of software requirement specifications (SRS) using Natural Language Processing (NLP) methods has been an important study area in the software engineering field in recent years. 
Especially thanks to the advances brought by deep learning and transfer learning approaches in NLP, SRS data can be utilized for various learning tasks more easily. 
In this study, we employ a three-stage domain-adaptive fine-tuning approach for three prediction tasks regarding software requirements, which improve the model robustness on a real distribution shift. 
The multi-class classification tasks involve predicting the \textit{type}, \textit{priority} and \textit{severity} of the requirement texts specified by the users.
We compare our results with strong classification baselines such as word embedding pooling and Sentence BERT, and show that the adaptive fine-tuning leads to performance improvements across the tasks. 
We find that an adaptively fine-tuned model can be specialized to particular data distribution, which is able to generate accurate results and learns from abundantly available textual data in software engineering task management systems.
}

% \keywords{Software Requirement Specification (SRS), Conflict Detection, Semantic Similarity, Named Entity Recognition (NER)}
%%%%%%%%%%%%%%%%%%%%%%%%%%%%%%%%%%%%%%%%%%%%%%%%%%%%%%%%%%%%%%%%%%%%%%%%%%%%%%%%%

\maketitle

%%%%%%%%%%%%%%%%%%%%%%%%%%%%%%%%%%%%%%%%%%%%%%%%%%%%%%%%%%%%%%%%%%%%%%%%%%%%%%%%%
\section{Introduction}
%%%%%%%%%%%%%%%%%%%%%%%%%%%%%%%%%%%%%%%%%%%%%%%%%%%%%%%%%%%%%%%%%%%%%%%%%%%%%%%%%

% Sizin diğer makaleden aldım
% (diger makale-START)\\
%  What is SRS
Software Requirements Specifications (SRS) provide detailed information on the functionality and features of a software project. 
They also help evaluate compliance with the standards and requirements laid down at the project initiation. 
SRS data is usually created in natural language and serves the purpose of enabling communication between the teams collaborating on the project. 
SRS language is subject to a certain level of subjectivity that might lead to misinterpretation, contradiction, ambiguity and redundancy. Automation brought by Natural Language Processing (NLP) methods can help avoid these issues in an effective manner.

NLP methods can be used to facilitate different tasks in requirement engineering including classification and extraction of requirements, which can lead to substantial savings in terms of time and effort spent on these tasks. 
The research in this field has benefited from recent developments in NLP for a variety of tasks such as extractive summarization, tagging, and entity recognition~\citep{kicibert}. 
These recent developments in NLP methodology can be attributed to transfer learning that has become the state-of-the-art approach for generic language modeling~\citep{ruder2019transfer}. Most transfer learning approaches rely on sequential learning, which has been a common practice for NLP due to its simplicity. 
Many NLP tasks are gradually achieved in sequential order as the generalization capacity of a language model increases with sequential learning that enables leveraging the information obtained from the previous task to achieve better performance on further tasks. 

Most studies in the literature rely on a two-stage transfer learning regimen: pre-training and fine-tuning. 
One main problem with the two-stage regimen is that the distribution of source data may differ from the target data. 
As such, it can be highly beneficial to adapt a model from a training distribution to the different distribution of the target task. 
Adaptive fine-tuning can provide a remedy for this problem by utilizing the unlabeled textual data in the target domain.
Accordingly, the two-stage regimen can be extended to three or more stages where additional training can be performed after the pre-training step. 
Such additional training is also known as adaptive pre-training (APT), pre-finetuning (PFT), and continual pre-training (CPT). 

\paragraph{Research goals and scope} 
In this study, we focus on leveraging transfer learning methods to improve text classification performance over SRS data. 
Specifically, we design an adaptive fine-tuning framework that makes use of abundantly available unlabeled data in the target domain.
We conduct our analysis with the SRS data obtained from IBM Rational DOORS Next Generation product~\citep{doors}.
This product serves as a requirement management tool for optimizing collaboration and communication between software development teams to improve the information flow within a project. 
DOORS dataset is used for three different classification tasks, namely, predicting \textit{Priority}, \textit{Severity}, and \textit{Type} of a requirement.
% (diger makale-END)

\paragraph{Contribution}
Standard transfer learning is typically performed in two stages. 
The first stage includes learning from source data, while the second stage is fine-tuning source knowledge for the solution of the main target task. 
However, the assumption that source and target share a similar distribution may not hold in real-world problems since the distribution of the target can differ from that of source data over time or across domains, which is called Out-of-Distribution (OOD) problem~\citep{hendrycks2020pretrained}. 
The solution is to adapt a model from the source distribution to the distribution of the target task.  
In this study, we show how to apply adaptation and leverage the three-stage domain-adaptive fine-tuning for three multi-class classification tasks over a software requirement dataset, DOORS.
The main contributions of our study are summarized as follows:
% Our study provides a comparative analysis for text classification over SRS documents on three different classification tasks. 

\begin{itemize}\setlength\itemsep{0.3em}
    \item We create a strong baseline for our analysis by utilizing word embeddings and Sentence BERT as feature extraction models where the model parameters are frozen and given as input to the linear classifier.
    % \item 
    % We also utilized Sentence BERT as feature extraction model
    % \item 
    Additionally, we utilize two- and three-stage fine-tuning mechanisms for the classification task where the base (e.g., BERT-base and RoBERTa-base), large (e.g., BERT-large and RoBERTa-large) and distilled versions of the Transformer models are leveraged.
    
    % \item Even though fine-tuning mechanisms are typically quite effective, the differences in the distribution of source and target data have significant effects on the fine-tuning performance. In this regard, 
    \item We make use of the additional unlabeled in-domain data to further pre-train Transformer model checkpoints on in-domain data. 
    It is an additional \textit{adaptive fine-tuning} step for improving text classification performance. 
    These adaptively trained models are then fine-tuned again for three downstream tasks.
    Designing an adaptive fine-tuning mechanism that provides a performance boost constitutes the main novelty of this work along with the novel application problem.
    
    \item We conduct an extensive empirical study, and provide a detailed analysis of the performances of various models for the SRS datasets.
    % \item 
    These analyses contribute to the understanding of the strengths and limitations of state-of-the-art text classification methods for important practical problems in software engineering.
\end{itemize}

\paragraph{Organization of the paper}
The rest of the paper is organized as follows.
In Section~\ref{sec:litreview}, we briefly review the most relevant studies to our work, with a special focus on NLP-based automation in software engineering and methodological developments in transfer learning.
We provide a discussion on the employed methods in Section~\ref{sec:methodology}, along with a discussion on our dataset and a review of Transformer models and domain adaptation.
In Section~\ref{sec:results}, we discuss the results from our numerical study, and we conclude the paper in Section~\ref{sec:conclusion} with a summary of our findings, and a discussion on study limitations and future research directions.

%%%%%%%%%%%%%%%%%%%%%%%%%%%%%%%%%%%%%%%%%
\section{Literature review}\label{sec:litreview}
%%%%%%%%%%%%%%%%%%%%%%%%%%%%%%%%%%%%%%%%%
% \tcolR{TO BE ADDED!!  Buraya sizin son makaleden (Derya Kici'nin) ekledim. Bunu re-phrase ediyorum... Yeterli bilgiler var zaten }\\
%%%%%%%%%%%%%%%%%%%%%%%%%%%%%%%%%%%%%%%%%%%%%%%%%%%%%%%%%%%%%%%%

In the literature, SRS data have been analyzed using various text classification techniques such as document classification or sentiment analysis, to enhance the software development process via high-quality predictions.
In earlier studies, traditional machine learning models were successfully used for such tasks. 
Hussain et al.~\citep{hussain2007using} applied document classification in a traditional machine learning pipeline to detect ambiguities in the SRS documents. 
They trained decision trees to map the SRS document passages to ambiguous or unambiguous labels, which is a binary classification task. 
Zhang et al.~\citep{zhang2013extracting} focused on extracting problematic API designs using sentiment analysis.
They extracted useful information from online resources that contain huge amounts of unstructured data, such as bug reports and some online discussions. 
Hou et al.~\citep{hou2013content} used Naive Bayes models to categorize API discussions based on their content. 
% They applied Naive Bayes Method as classifier. 
There have been various machine learning applications over the software requirement datasets.
For instance, Asabadi et al.~\citep{asadabadi2020ambiguous} employed fuzzy set theory to identify ambiguous SRS statements by using an ambiguous terms list. Apart from machine learning methods, rule-based approaches have also been employed for the classification tasks in the software engineering domain such as requirement classification~\citep{singh2016rule}.
In recent years, deep learning architectures have been frequently employed for text analytics tasks in requirement engineering. Navarro et al.~\citep{navarro2017towards} applied Convolutional Neural Network (CNN) models to classify software requirements without using handcrafted features.
Onyeka et al.~\citep{onyeka2019identifying} utilized commonsense knowledge ontology for implicit requirements framework where a CNN-based auto-encoder identifies implicit requirements from tables and images in large SRS documents.

The most recent NLP-related works in the field of software engineering and requirement engineering have been shaped by transfer learning. 
The pre-trained language models (PLM), specifically the encoder part of the Transformers, have been the best models to take advantage of transfer learning. 
In this regard, Bidirectional Encoder Representations from Transformers (BERT)~\citep{bert} has proven to be useful, especially for text and token classification problems such as sentiment analysis and named entity recognition. 
Many variants of BERT have recently emerged and they are used in software analytics including the classification tasks involving SRS data~\citep{kicibert, hey2020norbert,sainani2020extracting}. 
Hey et al.~\citep{hey2020norbert} fine-tuned BERT on specific tasks where the model predicts if a requirement is non-functional or functional. 
% The authors utilized a fine-tuning strategy to classify functional requirements based on the implied concerns. 
In another study, Sainani et al.~\citep{sainani2020extracting} employed BERT in a different way such that they first extract requirements from large software engineering contracts and then classify those. 
Their approach is based on the idea that business contracts could help in the identification of high-level requirements in order to improve the performance of software engineering projects.
Kici et al.~\citep{kicibert} fine-tuned various PLMs for different SRS tasks. 
They used different BERT models and the variants for three different datasets to test out the generalizability of their results.

All these aforementioned Transformer models have been implemented using a two-stage straightforward fine-tuning process. Specifically, they solved downstream tasks without domain-specific adaptation just by training the pre-trained model checkpoints with labeled data. 
However, it is possible to increase the transfer learning capacity by adaptive fine-tuning using unlabeled data that is easy to obtain for most problems. Adaptive fine-tuning is a domain adaptation technique to find a way of applying a pre-trained model trained on general-purpose source data to a different target domain~\citep{ramponi2020neural}. However, the source and target domain can have different distributions such as vocabulary distribution. Some studies~\citep{Gururangan2020, diao2021taming} showed that adaptation takes care of the effects of the mismatch between the previous source distribution and the target distribution. 
The domain-specific models such as BioBERT~\citep{Lee2019}, SciBERT~\citep{beltagy2019scibert}, HateBERT~\citep{Caselli2020}, ClinicalBERT~\citep{Alsentzer2019}, NetBERT~\citep{Louis2020}, MathBERT~\citep{Peng2021}, News~\citep{Gururangan2020} and GraphCodeBERT~\citep{Guo2020} have started to be frequently observed in the literature. 
All these models or checkpoints try to build a better model that can work in cases where the distribution of the target domain is different from that of the source domain. 
Some studies investigated the benefit of masking rate in Transformer architecture. 
The selection of 15\% has been mostly seen as an efficient default rate during pre-training of MLMs~\citep{Clark2020}. 
On the other hand, in another study, \citet{Wettig2022} suggested that masking around 40\% of input tokens instead of 15\% can show a better downstream task performance.

To the best of our knowledge, there is no other study that focused on software requirement domain adaptation. 
In our study, we addressed this research gap in the software engineering area and applied adaptive fine-tuning over software requirement data to improve the text classification performance for a specific task.

%%%%%%%%%%%%%%%%%%%%%%%%%%%%%%%%%%%%%%%%%
\section{Methodology}\label{sec:methodology}
%%%%%%%%%%%%%%%%%%%%%%%%%%%%%%%%%%%%%%%%%
In deep learning models, the trained parameters of the neural network architecture can be stored for future use based on two types of transfer: feature extraction and fine-tuning. 
The parameters obtained in the former are fixed as in word embeddings and not subject to gradient descent, whereas the parameters obtained in the latter are subject to change as in the BERT model.
In this section, we explain how we employed feature extraction briefly discuss the standard two-stage training process, and elaborate on adaptive fine-tuning framework.

\subsection{DOORS dataset}
Our dataset consists of software requirement specifications collected using IBM's Dynamic Object-Oriented Requirements System Next Generation (DOORS) product. 
The original dataset contains 83,837 documents. % and 213 features. 
We consider the \textit{Summary} of the requirements specifications to classify for three separate categories: \textit{Priority}, \textit{Severity}, and \textit{Type}. 
In this regard, the DOORS dataset can be considered as a task management dataset, and differs from other SRS datasets in the literature (e.g., see NFR-PROMISE~\citep{hey2020norbert}), which are typically used for requirement type classification (e.g., functional vs nonfunctional).
Table~\ref{tab:dataSample} provides data instances from the DOORS dataset.

\setlength{\tabcolsep}{6pt} % adjust column separation in table
\renewcommand{\arraystretch}{1.35} % adjust row separation in table
\begin{table}[!ht]
\centering
\caption{DOORS data samples}\label{tab:dataSample}
% \caption{\tcolR{DOORS data sample -- CHANGE IF YOU FIND BETTER INSTANCES}}
\resizebox{1.01\linewidth}{!}{
\begin{tabular}{|l|l|l|l|}
\hline
\textbf{Summary} & \textbf{Type} & \textbf{Priority} & \textbf{Severity} \\
\hline
Default selection by creating a new link must be MODULE and NOT FOLDER & Enhancement & Medium & Major \\
\hline
Provide easy mechanism to update server and test server rename & Task & Medium & Normal\\
\hline
As a developer, I should be able to implement support for large lists of Longs in view queries & Story & Unassigned & Normal\\
\hline
Export artifact to a PDF file is not working & Defect & High & Major\\
\hline
All Projects label should be updated in the component selection page & Enhancement & Unassigned & Minor\\
\hline
Investigate potential corrupt data involving wrapper resources on self host & Issue & Low & Normal\\
\hline
\end{tabular}
}
\end{table}

Table~\ref{tab:eda1} shows the distribution of the classes for each category label.
There exist four classes in the \textit{Priority} (\textit{unassigned, high, medium, and low}), and six classes in \textit{Severity} (\textit{normal, major, minor, blocker, critical, and undecided}). 
\textit{Type} category originally consisted of 20 different classes. 
With some merging and pre-processing operations, we obtained seven classes for the \textit{Type} category. 
In addition, we eliminated `nan' values, which correspond to missing values both for \textit{Severity} and \textit{Priority} categories.
We note that while the Type category's classes are somewhat fairly distributed, the others show severe data imbalance. 

\begin{table}[!ht]
\centering
\caption{Class distribution for each category label}
\label{tab:eda1}
    \subfloat[Priority \label{tab:eda1_priority}]{
        \resizebox{0.325\textwidth}{!}{
        \begin{tabular}{|l|r|r|}
        \hline
        \textbf{Class} & \textbf{Count} & \textbf{Perc. (\%)} \\ \hline
        Unassigned & 11,443 & 68.97 \\ \hline
        High & 2,677 & 16.13 \\ \hline
        Medium & 1,990 & 11.99 \\ \hline
        Low & 480 & 2.89 \\ \hline
        \end{tabular}
        }
    }%\hfill
    \subfloat[Severity \label{tab:eda1_severity}] {
    \resizebox{0.325\textwidth}{!}{
        \begin{tabular}{|l|r|r|}
        \hline
        \textbf{Class} & \textbf{Count} & \textbf{Perc. (\%)} \\
        \hline
        Normal & 14,348 & 86.48 \\
        \hline
        Major & 1,127 & 6.79 \\
        \hline
        Undecided & 478 & 2.88 \\
        \hline
        Minor & 284 & 1.71 \\
        \hline
        Critical & 245 & 1.47 \\
        \hline
        Blocker & 108 & 0.65 \\
        \hline
        \end{tabular}
        }
    }%\hfill
    \subfloat[Type \label{tab:eda1_type}] {
    \resizebox{0.325\textwidth}{!}{
    \begin{tabular}{|l|r|r|}
    \hline
    \textbf{Class} & \textbf{Count} & \textbf{Perc. (\%)} \\
    \hline
    Enhancement & 5,026 & 30.29 \\
    \hline
    Story & 4,565 & 27.50 \\
    \hline
    Maintenance & 2,180 & 13.14 \\
    \hline
    Other & 1,774 & 10.69 \\
    \hline
    Test Task & 1,615 & 9.73 \\
    \hline
    Plan Item & 885 & 5.33 \\
    \hline
    JUnit & 545 & 3.28 \\
    \hline
    \end{tabular}
    }
    }
\end{table}

\begin{figure}[!ht]
    \centering
\includegraphics[width=0.925\textwidth]{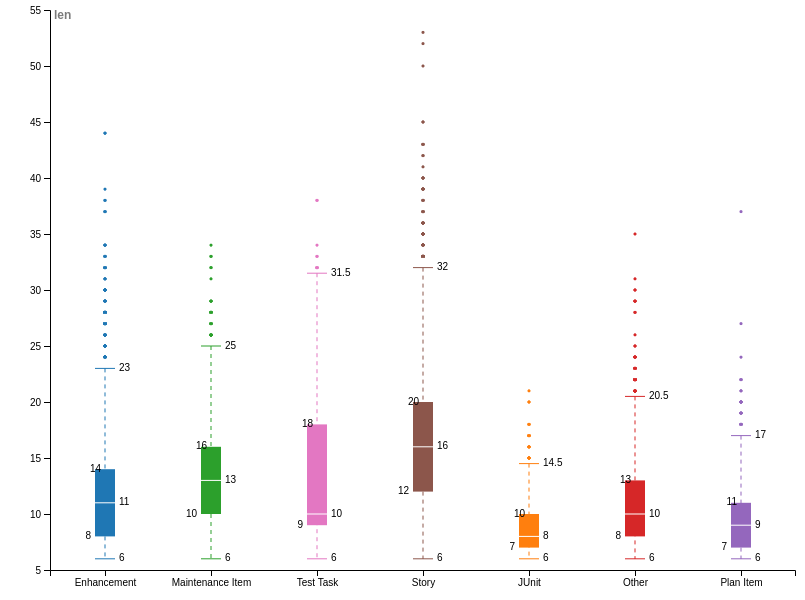}
    \caption{Text length box plot across Type label classes in the DOORS dataset}
    \label{fig:eda_boxplot_type}
\end{figure}

Having applied the pre-processing steps, the total number of remaining instances is 16,590, where the text length is on average 13.1 words with a median value of 12.
We provide the box plot of text lengths by class for the Type label as a representative case in Figure~\ref{fig:eda_boxplot_type}.
We find that text length distribution over the classes provides some information for the Type label (e.g., Story class has the longest text and JUnit has the shortest).
However, text length distribution across the classes is more uniform for Priority and Severity labels.

%%%%%%%%%%%%%%%%%%%%%%%%%%%%%%%%%%%%%%%%%%%%%%%%%%%%%%%%%%%%%%%%%%%%%%%%%%
\subsection{Feature extraction methods}
%%%%%%%%%%%%%%%%%%%%%%%%%%%%%%%%%%%%%%%%%%%%%%%%%%%%%%%%%%%%%%%%%%%%%%%%%%
Learning text representations through averaging fixed vectors of the words has been found to be a simple but effective method in previous works. 
The word embedding techniques such as FastText~\citep{fasttext}, Glove~\citep{glove} and Word2vec~\citep{word2vec} can provide strong static word vectors that are kept for a different purpose in a generic machine learning pipeline, called \textit{feature extraction}. 
That is, word vectors (or document vectors) are kept frozen, and provided as input to a linear classifier or a clustering algorithm. 

Such document representations based on averaging static word vectors have become surprisingly strong baselines. 
In some cases, these approaches can achieve higher performance than more complex models~\citep{wieting2015towards}. 
Moreover, they provide a basis for whether there is a pattern in the data. 
For instance, Cer et al.~\citep{use} proposed an averaging-based model, called the DAN network, and showed that their model can achieve similar performance to those of more complex architectures.
On the other hand, the averaging method has certain limitations as well, since it is based on frozen word embeddings. 
Other than averaging, different sentence-level approaches have been proposed, which are based on feature extraction for text representation such as Doc2vec \citep{doc2vec}, Sentence BERT \citep{sbert}, and Skip Thought \citep{kiros2015skip}. 

% \subsection{Baseline models}
% We designed strong baseline models to see how successful our fine-tuned models are. 

In our experiments, we employ two feature extraction methods as strong baselines: Word Embedding Pooling and Sentence BERT. 
We describe the settings for these two approaches as follows.
% to building a good benchmark in addition to a simple majority-based model as follows:

\begin{itemize}\setlength\itemsep{0.3em} 
% \item \textit{The majority-based classifier}: It is one of the simplest classifiers in machine learning. It predicts as the class of a new sentence the most frequent class label. We may measure the contribution of other models by taking it as a reference.
\item \textit{Word Embedding Pooling}: In Word Embedding Pooling, we simply take an average of word embeddings in a sentence where FastText and Glove pre-trained vectors are obtained and concatenated. 
The length of the final vector is 400. 
A linear layer is added on top of it, and the model is trained in a supervised pipeline. 

\item \textit{SBERT}: Sentence BERT embeddings have been found to be a highly efficient way of text representation, especially for semantic problems such as sentence similarity or semantic search. 
The main motivation behind SBERT approach is that a BERT model is not suitable for the standalone usage of word or sentence vectors since it requires end-to-end fine-tuning for a downstream task. Based on Siamese network structures, SBERT can produce semantically meaningful and independent embeddings of the sentences, making it a suitable candidate for a baseline method. 
% Therefore, SBERT can be considered as a good candidate for a baseline. 
\end{itemize}

%%%%%%%%%%%%%%%%%%%%%%%%%%%%%%%%%%%%%%%%%%%%%%%%%%%%%%%%%%%%%%%%%%%%%%%%%%
\subsection{Two-stage training: Pre-training and fine-tuning}\label{sec:twostageTraining}
%%%%%%%%%%%%%%%%%%%%%%%%%%%%%%%%%%%%%%%%%%%%%%%%%%%%%%%%%%%%%%%%%%%%%%%%%%
To fine-tune a pre-trained source model with the parameters $\Theta_{S}$ to a different target task, task-specific parameters $\Theta_{T}$ (or layers) are designed and added to $\Theta_{S}$.
For feature extraction, as applied mostly with word embeddings, $\Theta_{S}$ is frozen, and only $\Theta_{T}$ parameters are updated for a given task. 
For fine-tuning, as applied in the ELMo~\citep{elmo} and the BERT models~\citep{bert}, the entire set of parameters, $\Theta_{S} \cup \Theta_{T}$, is updated, mostly in an end-to-end fashion. 
During fine-tuning, the learning rate is typically set to a smaller value than the one in pre-training settings in order not to update the main model too frequently. 
This is due to the observation that most syntactic and semantic information has been already encoded in the pre-trained models. 
On the other hand, an important issue is the difference in the distribution of the data between the target and the source. 
For instance, a word that does not appear in the pre-training phase but appears in the fine-tuning phase leads to an out-of-vocabulary problem~\citep{hendrycks2020pretrained}. 

ELMo~\citep{elmo} and Transformer models~\citep{attention} have made great contributions to natural language understanding and generation problems. 
Their success can be attributed to the fact that these architectures encode information by distributing it into layers in deep networks, rather than encoding it in a simple one-dimensional vector. 
They have been successfully utilized in the fine-tuning phase, and are capable of transferring the knowledge obtained at previous training phases.
Recently, Transformer architectures~\citep{attention} have shown promising results for many tasks due to utilizing the self-attention mechanism. 
One of the most important advantages of Transformer architectures is that they can create reliable pre-trained models in a parallelizable architecture, making transfer learning faster and more effective. 
In addition to being very suitable for domain adaptation, the Transformer models are also highly robust to OOD samples~\citep{hendrycks2020pretrained}.

% how fine-tuning works 
Contrary to word embedding-based averaging models, a Transformer model is trained in an end-to-end fashion. 
A thin layer, $\Theta_{T}$, is added on top of a pre-trained source model, $\Theta_{S}$, and the entire architecture, $\Theta_{S} \cup \Theta_{T}$, is trained as a whole, called fine-tuning. 
In some cases, it would be computationally expensive to fine-tune the entire architecture. 
Some approaches in this line have been developed by utilizing adapters~\citep{adapter,rebuffi2017learning}, which is based on the idea that training the entire architecture is avoided by adding simple trainable adapters between layers instead. Another reason for using the adapters is that the models can suffer from catastrophic forgetting, which means information learned during previous stages can be lost when adding new tasks.
In our analysis, we took this negative transfer notion into account by observing the losses and the performance, which is important for not disrupting our pre-trained models and also not suffering from catastrophic forgetting and negative transfer.

The fine-tuning strategy has been shown to perform better than its static feature extraction counterpart for various problem instances. 
In the numerical study, we consider widely used models and their checkpoints for fine-tuning, which are briefly summarized below. 
\begin{itemize}\setlength\itemsep{0.3em}
    \item \textbf{BERT checkpoints:} BERT is the most well-known encoder using both Masked Language Model (MLM) and Next Sentence Prediction (NSP) objectives. The base model, \textit{BERT-base-uncased}, has a hidden size of 768, which corresponds to 12 heads and a head embedding size of 64. 
    The number of layers in the model is 12. 
    The large counterpart model, \textit{BERT-large-uncased}, has a hidden size of 1024 with 24 layers and 16 attention heads. 
    The original learning rate used for BERT checkpoints is 1e-4. 
    When it is fine-tuned, a smaller learning rate (e.g., 1e-5 or 2e-5) is selected in order not to disrupt the learning during pre-training. 
    
    \item \textbf{DistilBERT checkpoints:} We use \textit{distilbert-base-uncased} checkpoint of the DistilBERT~\citep{distilbert}, which is a small and light model trained by distilling BERT-base checkpoint. It is pre-trained on the same data used to pre-train BERT-base model. The corpus consists of the Toronto Book Corpus and full English Wikipedia using distillation with the supervision of the \textit{BERT-base-uncased} version. The model has 6 layers, a hidden size of 768 with 12 heads, and 66M parameters.

    \item \textbf{RoBERTa checkpoints}: RoBERTa (Robustly Optimized BERT pre-training Approach)~\citep{roberta} is a well-known BERT re-implementation. 
    RoBERTa training provided many more improvements in terms of training strategies than modifying the architectural design. 
    For instance, NSP training objective is removed, and the static masking is replaced by the strategy of dynamically changing the masking patterns.
    In our analysis, Adam optimizer is used during pre-training as in other BERT checkpoints with a learning rate of 6e-4.
\end{itemize}

%%%%%%%%%%%%%%%%%%%%%%%%%%%%%%%%%%%%%%%%
\subsection{Adaptive fine-tuning}
%%%%%%%%%%%%%%%%%%%%%%%%%%%%%%%%%%%%%%%%
Even though fine-tuning approach used for the Transformer architectures typically performs well, the differences in the distribution of source and target data typically have a significant impact on the effectiveness of fine-tuning~\citep{ruder2019transfer}. 
If the source and target datasets are substantially different from each other, fine-tuning may face difficulty in learning. 
Typically, NLP model evaluations rely on the assumption that the source (train) and target (test) instances are independent and identically distributed, which does not hold in most cases in real-world applications.  
Such discrepancy may appear since the target dataset hardly characterizes the entire distribution, and the target distribution usually changes over time~\citep{torralba2011unbiased,quinonero2009dataset}. 
Previous studies showed that pre-trained Transformers are the most robust architecture on real-world distribution shift and have more generalization capacity than other architectures, however, there still remains room for improvement~\citep{hendrycks2020pretrained}. 

\begin{figure}[!ht]
    \centering
\includegraphics[width=0.81\textwidth]{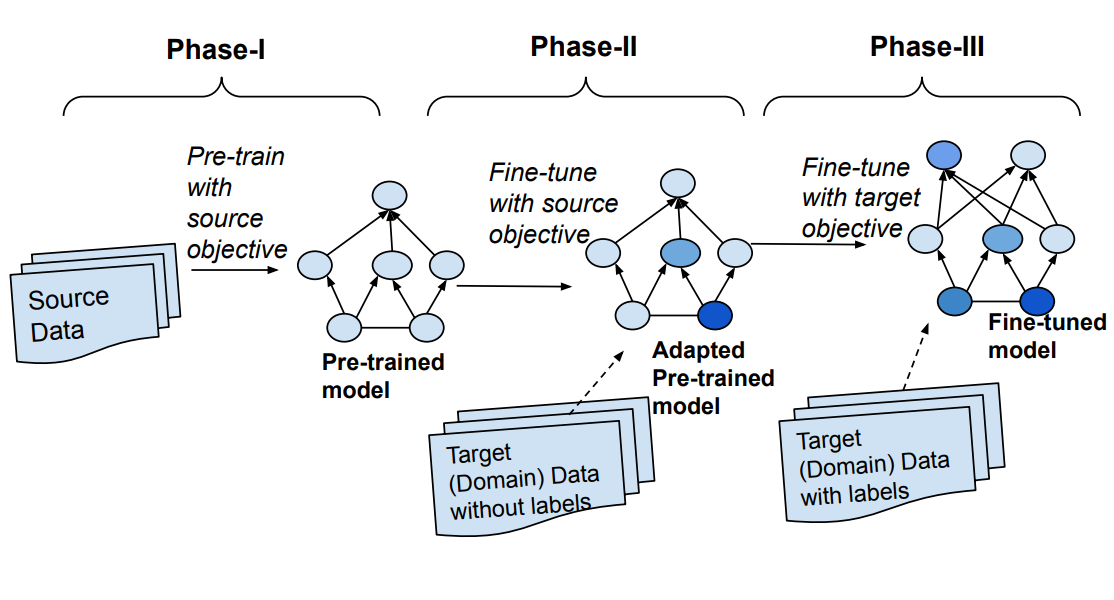}
    \caption{Adaptive fine-tuning framework}
    \label{fig:adaptive_fine_tuning}
\end{figure}

In the literature, several strategies were proposed to adapt a pre-trained model to a target domain~\citep{ruder2019transfer}. 
In our implementations, we keep training the pre-trained models using some additional in-domain data, with the expectation that specializing the model to target data would improve the downstream task performance.
That is, an already pre-trained model is continually trained with the pre-training objective on target data (i.e., $J_{S}= J_{A}$) that is expected to be closer to the target distribution. 
Finally, we end up with another version of the pre-trained model, $f(\hat{\theta}_{S})$, that still requires fine-tuning to a downstream task. 
Figure~\ref{fig:adaptive_fine_tuning} summarizes our adaptive fine-tuning framework.
% We summarize this adaptive process in Figure~\ref{fig:adaptive_fine_tuning}.

Adaptive fine-tuning consistently helps with distribution shifts. 
Therefore, just before the fine-tuning phase, the model can be further trained with the target dataset. 
In other words, we hold three phases as depicted in Figure~\ref{fig:adaptive_fine_tuning}. 
The first phase includes a pre-training process to learn the parameters $\theta_{S}$ with source objective $J_{S}$ where we do not need any labeled data. 
The objective function is typically the masked language model. 
In fact, there are already many checkpoints trained for this purpose in the literature. Therefore we did not have to pre-train a model from scratch. 
These pre-trained models have been trained on a large variety of data and have a sufficient depth of architecture and a number of parameters as discussed in Section~\ref{sec:twostageTraining}. 
% In our numerical analysis, we utilize many different pre-trained checkpoints.

In the second phase, the pre-trained model is kept training, again with the same source objective $J_{S}= J_{A}$ but with target datasets for adaptation. 
It is important to note that other kinds of auxiliary objective functions can be utilized to improve the adaptation. 
At this phase, we do not change the Transformer model architecture and do not add any additional parameters. 
Furthermore, in the third phase, new parameters $\theta_{T}$ are added to the architecture. 
They are usually found in the new layer, which is placed on top of the last layer.
The new architecture is then trained with a target objective $J_{T}$ and target labeled data in an end-to-end fashion. 
During this phase, we set the learning rate to a smaller value than one in the pre-training and adaptive fine-tuning phases in order not to disrupt the main model significantly, as the model learns the most important syntactic and semantic information by this stage.
% This is because the most important syntactic and semantic information has been already learned and encoded in the model. 
% We also evaluated fine-tuning learning rates at the last stage in terms of their loss performance as shown in Figure~\ref{fig:adative_fine_tuning_loss}.

%%%%%%%%%%%%%%%%%%%%%%%%%%%%%%%%%%%%%%%%
%%%%%%%%%%%%%%%%%%%%%%%%%%%%%%%%%%%%%%%%
\subsection{Experimental settings}
%%%%%%%%%%%%%%%%%%%%%%%%%%%%%%%%%%%%%%%%
%%%%%%%%%%%%%%%%%%%%%%%%%%%%%%%%%%%%%%%%
Table~\ref{tab:hyperparameters} lists the hyperparameters used for different Transformer models. 
Specifically, we experiment with different BERT checkpoints with varying sizes. 
The learning rate (\textit{lr}) and activation function parameters are determined via hyperparameter tuning experiments.

\begin{table}[!ht]
\centering
\caption{Transformer model hyperparameters}\label{tab:hyperparameters}
\resizebox{0.99\textwidth}{!}{
\begin{tabular}{|l|r|r|r|r|r|c|}
\hline
\textbf{Check point} & \textit{\textbf{Layers}} & \textit{\textbf{Hidden units}} & \textit{\textbf{Heads}} & \textit{\textbf{Total Params}} & \textit{\textbf{lr}} & \textit{\textbf{activation}} \\ 
\hline
DistilBERT& 6 & 768 & 12 & 66M & 5e-4 & gelu \\ 
\hline
BERT-base & 12 & 768 & 12 & 110M & 1e-4 & gelu \\ 
\hline
BERT-large& 24 & 1024 & 16 & 340M & 1e-4 & gelu \\ 
\hline
RoBERTa-base & 12 & 768 & 12 & 125M & 6e-4 & gelu \\ 
\hline
RoBERTa-large & 24 & 1024 & 16 & 355M & 4e-4 & gelu \\ 
\hline
\end{tabular}
}
\end{table}

%%%%%%%%%%%%%%%%%%%%%%%%%%%%%%%%%%%%%%%%%%%%%
% \subsection{Adaptive fine-tuning settings}
%%%%%%%%%%%%%%%%%%%%%%%%%%%%%%%%%%%%%%%%%%%%%
In a typical supervised training pipeline, unlabeled data would be eliminated and the fine-tuning phase would be implemented on a labeled dataset. 
In the DOORS dataset, there are more than 80,000 requirement texts, however, the majority of those are unlabeled. 
Accordingly, we employ all the requirement texts from the entire dataset for the adaptive fine-tuning process by using MLM as the unsupervised learning objective. 
The tokens are randomly masked in the input with a probability of 0.15 to implement MLM. 
This process yields a new adapted pre-trained model with the updated parameters, $\hat{\theta}_{S}$.

For the rest of the training procedure, we follow the common practice to apply hyper-parameter selection. 
Dataset is divided into three sets: training, validation, and test. The learning rate is kept around 1e-4 as in other pre-training hyperparameter settings such as BERT checkpoints. 
The model is trained up to 30 epochs, using \textit{AdamW} optimizer~\citep{adamW}. 
The best model found during the training is loaded at the end. 
Then fine-tuning phase is applied as the last phase. 

%%%%%%%%%%%%%%%%%%%%%%%%%%%%%%%%%%%%%%%%%%%%%
% \subsection{Fine-tuning settings}
%%%%%%%%%%%%%%%%%%%%%%%%%%%%%%%%%%%%%%%%%%%%%
We also follow the common practice at the fine-tuning phase that comes after the adaptive phase. 
Specifically, we have tree classification targets \textit{Priority}, \textit{Severity}, and \textit{Type} as downstream tasks. 
For each task, we fine-tuned a pre-trained model, that is an adapted pre-trained model obtained at stage two, up to 3 epochs. 
We observe that keeping the learning rate value very small at this stage is highly important so as not to spoil the adapted language model. 

%%%%%%%%%%%%%%%%%%%%%%%%%%%%%%%%%%%%%%%%%%%%%%%%%%%%%%%%%%%%%%%%
%%%%%%%%%%%%%%%%%%%%%%%%%%%%%%%%%%%%%%%%%%%%%%%%%%%%%%%%%%%%%%%%
% \section{Results and Discussion}\label{sec:results}
\section{Numerical Study}\label{sec:results}
%%%%%%%%%%%%%%%%%%%%%%%%%%%%%%%%%%%%%%%%%%%%%%%%%%%%%%%%%%%%%%%%
%%%%%%%%%%%%%%%%%%%%%%%%%%%%%%%%%%%%%%%%%%%%%%%%%%%%%%%%%%%%%%%%
In this section, we provide the results from our detailed numerical study. We first show representative results from the hyperparameter tuning experiments.
Then, we present the results from our comparative analysis with different baselines and Transformers, along with their adaptively fine-tuned variants.
We next examine the performance of the best-performing model using the class-specific outcomes.
Lastly, we provide a discussion on model predictions and elaborate on potential causes for misclassifications.

%%%%%%%%%%%%%%%%%%%%%%%%%%%%%%%%%%%%%%%%%%%%%%%%%%%%%%%%%%%%%%%%
\subsection{Hyperparameter tuning results}
%%%%%%%%%%%%%%%%%%%%%%%%%%%%%%%%%%%%%%%%%%%%%%%%%%%%%%%%%%%%%%%%

\begin{figure}[!ht]
    \centering
    \subfloat[Fine-tuning loss ($x$-axis: training steps) \label{fig:fine_tuning_loss}]{\includegraphics[width=0.685\textwidth]{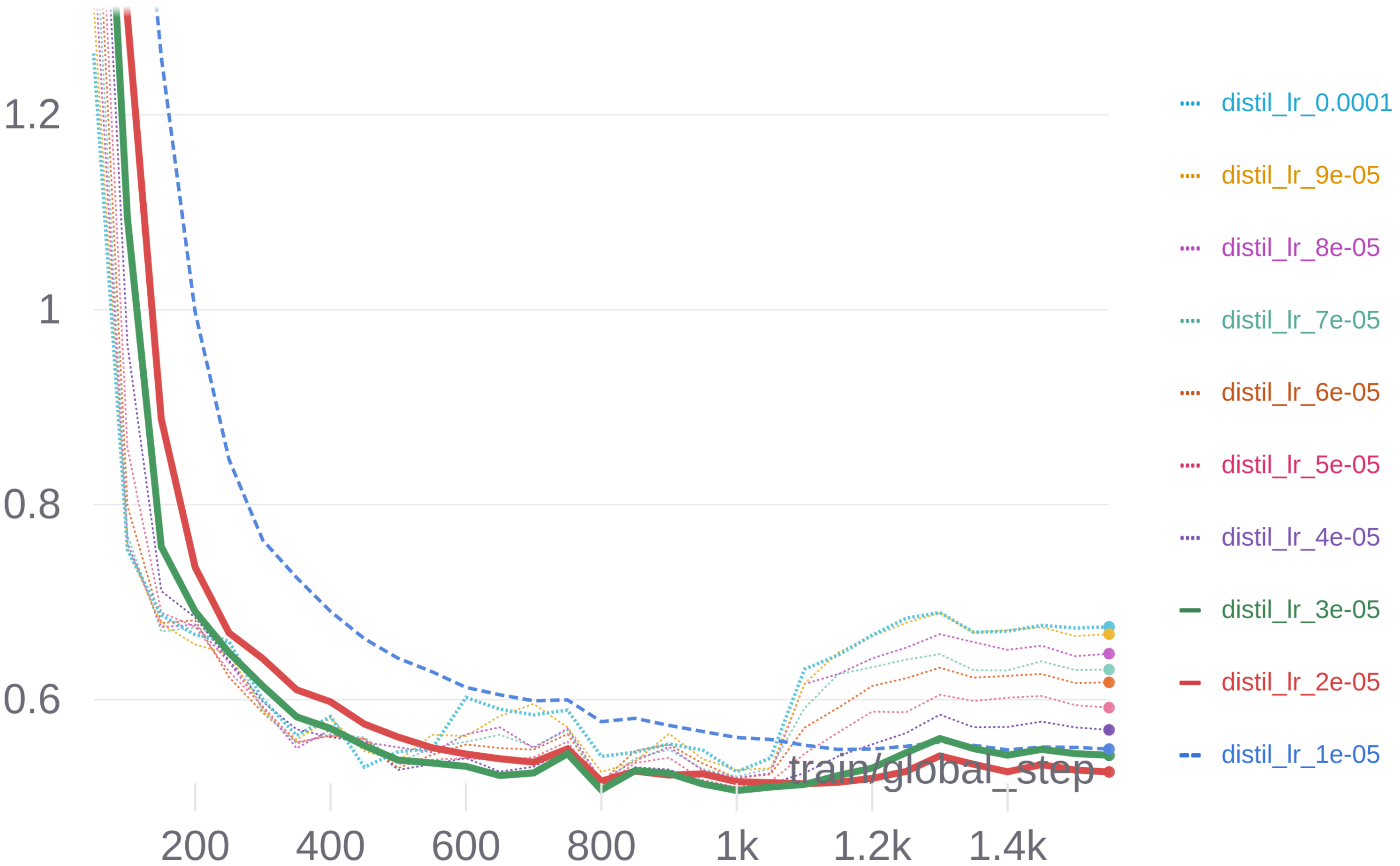}}\\
    \subfloat[Adaptive fine-tuning loss \label{fig:adative_fine_tuning_loss}]{\includegraphics[width=0.705\textwidth]{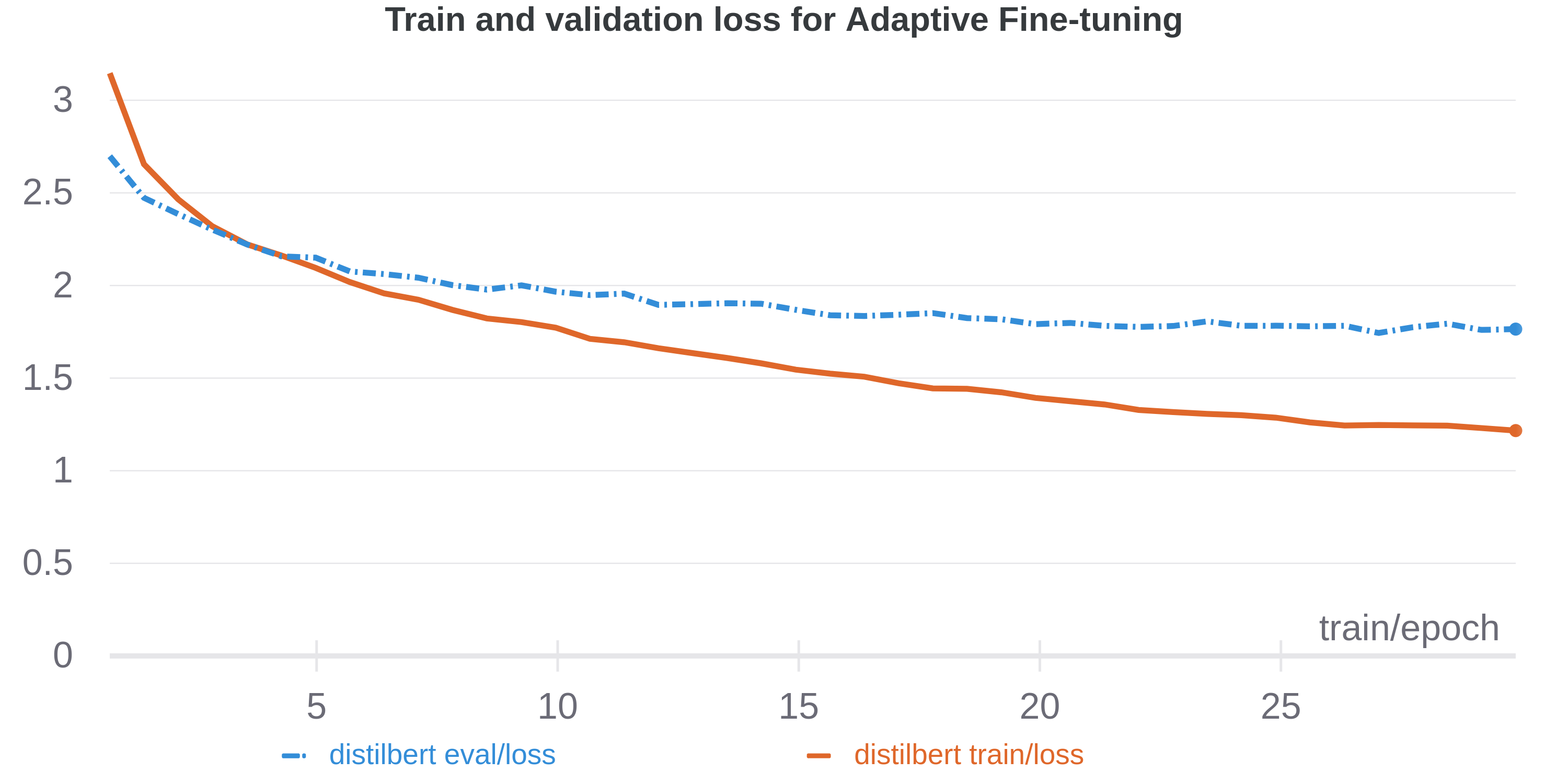}}
    \caption{Sample loss values observed during hyperparameter tuning for the selected pre-trained model, DistilBERT ($y$-axis: loss values).}
    \label{fig:lossValues}
\end{figure}

We performed hyperparameter tuning for various model parameters listed in Table~\ref{tab:hyperparameters}.
Figure~\ref{fig:fine_tuning_loss} illustrates the fine-tuning loss of the downstream task of Type prediction across different learning rates when fine-tuning a selected Transformer model (DistilBERT). 
The learning rate values between 1e-05 and 1e-04 have been evaluated. 
We observe that, for the lower rates such as 1e-05, even though training eventually converges to the ideal level, this can take a long time, e.g., up to 1500 steps in this case. 
As the learning rates get too high, the model training makes progress very quickly at first, but it never converges and settles down, which leads to divergence. 
When the learning rate is selected between 2e-05 and 3e-05 (illustrated as thick straight lines in the figure), we obtain satisfactory results and fine-tuning achieves convergence.

Figure~\ref{fig:adative_fine_tuning_loss} illustrates the training and validation loss for adaptive fine-tuning of a DistilBERT checkpoint.
We see that after 15 epochs, the reduction in validation loss almost stops and the model converges. 
We also observe that there is no variance and bias problem in the training process. 
% Similarly, Figure~\ref{fig:adative_fine_tuning_loss} shows the evaluation of fine-tuning learning rates at the last stage in terms of their loss performance.

%%%%%%%%%%%%%%%%%%%%%%%%%%%%%%%%%%%%%%%%%%%%%%%%%%%%%%%%%%%%%%%%
\subsection{Comparative performance analysis results}
%%%%%%%%%%%%%%%%%%%%%%%%%%%%%%%%%%%%%%%%%%%%%%%%%%%%%%%%%%%%%%%%
We report the model performances in terms of Accuracy (ACC), F1, and weighted F1 (w-F1)) as in Table~\ref{tab:perf_comparison_table}.
Three baseline models are employed, namely majority class classifier, Word Embeddings (WE) and SBERT.
The rest of the table contains the performance of the Transformers models. 
For easier comparison, we list the results of two-stage (pre-training and fine-tuning) learning and three-stage adaptive learning back-to-back as shown in the table. 
The three-stage process is indicated with the term ``+adapted''.

\begin{table}[!ht]
\centering
\caption{Comparison of accuracy values for the classification models (*: adaptively fine-tuned version). }
\label{tab:perf_comparison_table}
\resizebox{0.999\textwidth}{!}{
\begin{tabular}{|l|l|l|l|} 
\hline
ACC / F1 /w-F1~      & \multicolumn{1}{l}{Priority}                     & \multicolumn{1}{l}{Severity}           & Type                                               \\ 
\hline
Majority Classifier  & 68.39 / 20.31 / 55.56                            & 86.11 / 15.42 / 76.69                  & 30.16 / 6.62 / 13.98                               \\ 
\hline
\multicolumn{4}{|l|}{\textbf{Feature Extraction Baselines }}                                                                                                          \\ 
\hline
WE(Glove + FastText) & 71.52 / 31.82 / 66.18                            & 86.60 / \textbf{27.76} / 81.74         & 75.92 / 73.81 / 75.10~                             \\ 
\hline
SBERT                & 70.85 / 32.79 / 67.88                            & 86.62 /\textbf{ 25.59} / 82.52~ ~      & 77.22 / 74.78 / 76.33                              \\ 
\hline
\multicolumn{4}{|l|}{\textbf{Transformer Models~}}                                                                                                                    \\ 
\hline
DistilBERT           & 72.09 / 33.20 / 65.34                            & 86.43~/ 15.76 / 79.84                  & 78.80 / 73.88 / 76.72                              \\ 
\hline
\textit{+ adapted}   & \textit{\textbf{72.54}~/ 33.21~/ 65.45}          & \textit{\textbf{86.61} / 16.9 / 80.89} & \textit{\textbf{81.15}~/~ 77.57 \textbf{/ 79.24}}  \\ 
\hline
Bert-base            & 71.89 / 32.51 / 65.0                             & 86.49 / 16.3 / 80.48                   & 77.24 / 70.87 / 74.62~                             \\ 
\hline
\textit{+ adapted}   & \textbf{72.63} / \textbf{33.8 }/ \textbf{65.87~} & 86.56 / \textbf{22.3}~/ \textbf{81.64} & 80.06 / 75.6~~/ 77.76                              \\ 
\hline
Bert-large           & 68.39 / 20.31 / 55.56~                           & 86.12 / 15.42 / 79.68                  & 79.48 / 75.91 / 78.13~                             \\ 
\hline
\textit{+adapted}    & 71.94 / 30.64 / 65.29                            & 86.11 / 15.4 / 79.70                   & 79.52 / 76.06 / 78.19                              \\ 
\hline
Roberta-base         & 71.33 / 31.04 / 62.11                            & 86.32 / 15.11 / 79.29                  & 80.12 / 77.11 / 78.29~                             \\ 
\hline
\textit{+adapted}    & 71.78 / 32.12 / 65.12~                           & 86.21 / 29.43 / 81.49                  & 80.45 / \textbf{78.71} / 79.15~                    \\ 
\hline
Roberta-large        & 68.81 / 30.12 / 56.17~                           & 86.03 / 15.71 / 79.98                  & 79.59 / 75.94 / 78.17~                             \\ 
\hline
\textit{+adapted}    & 72.02 / 32.77 / 65.11                            & 86.23 / 15.67 / 80.40                  & 79.94 / 76.16 / 78.18                              \\
\hline
\end{tabular}
}
\end{table}

These results show that the Transformer models, with or without adaptation, outperform the majority classifier baseline model. 
They also outperform word embedding pooling and SBERT baselines for \textit{priority} and \textit{type} tasks. 
This can be attributed to the generalization capacity of the Transformers and the benefit of adaptation. However, we observe contradictory results for the \textit{Severity} category. 
%We will address this later.

\begin{figure}[!ht]
    \centering
    \includegraphics[width=0.8\textwidth]{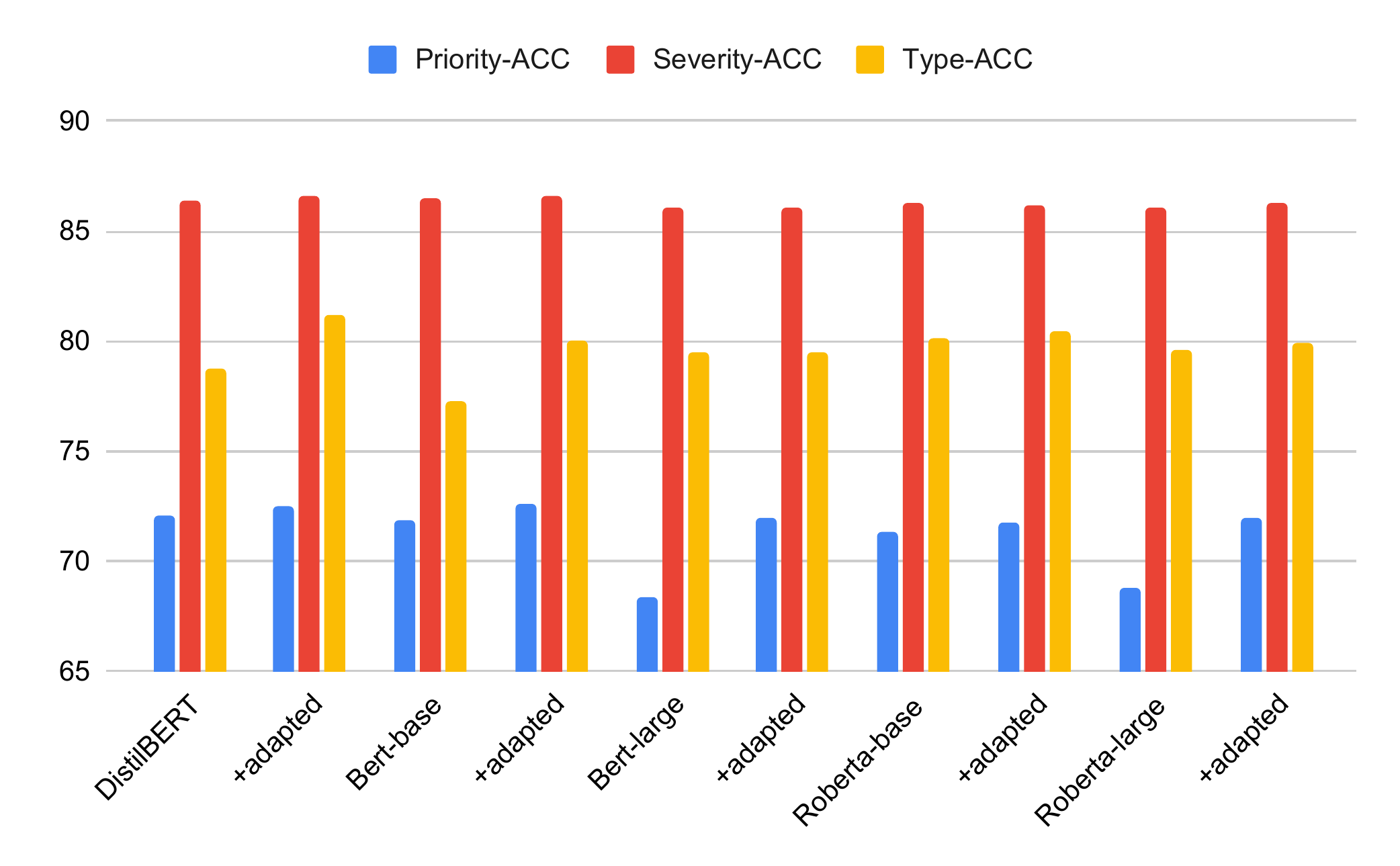}
    \caption{Performance comparison across adapted models based on accuracy values}
    \label{fig:perf_comparison_adapted}
\end{figure}

% When compared with baseline models, since Transformer models are very good at generalization capacity, all the base models with or without adaptation outperformed the baseline algorithms. 
Another observation is that the adaptation brought an improvement for all model settings. 
We could not see any negative transfer or performance loss. In addition, we examine which of these differences are statistically significant with a detailed analysis as some improvements have very low degree. 
We also note that all the adapted models achieve similar performance as shown in Figure~\ref{fig:perf_comparison_adapted}. 

We report the improvement obtained by adaptation as shown in Table~\ref{tab:p_value}. 
We list the improvement obtained by adaptation against its vanilla counterpart and majority classifier baselines. 
The $p$-value shows whether the difference is statistically significant. 
We employ the 5x2 cv F-test for this purpose, which is able to determine whether there is a significant difference between the performance of two classifiers~\citep{alpaydm1999combined}. This method employs 2-fold cross-validation 5 times for two classifiers to be compared. 
Thus, the values of the two models under the same conditions and on the same fold are comparable. 
If the $p$-value is less than $\alpha$ (0.05), the null hypothesis is rejected, which means that the two models are significantly different.

\begin{table}[!ht]
\centering
\caption{Improvement made by adaptation and its $p$-value as compared to vanilla Transformer model variants and the majority classifier baseline (reported as ``+Improvement(p-value)'')}
\label{tab:p_value}
\resizebox{0.99950\linewidth}{!}{
\begin{tabular}{|l|l|l|l|l|l|l|} 
% \multicolumn{7}{l}{\textbf{+Improvement (p-value)}}\\ 
\cline{1-7}
\cline{1-7}
\multicolumn{1}{l}{} & \multicolumn{3}{c|}{Vanilla Counterpart} & \multicolumn{3}{c}{Majority Classifier}                                      \\ 
\cline{2-7}
\multicolumn{1}{l}{\textbf{Priority}}  & \multicolumn{1}{|l}{Acc.} & \multicolumn{1}{|l}{F1} & \multicolumn{1}{|l}{w-F1} & \multicolumn{1}{|l|}{Acc.} & \multicolumn{1}{|l|}{F1} & \multicolumn{1}{|l|}{w-F1}  \\ 
\hline
Bert+adapted                                           &+0.73(0.15)           &+1.29(0.047)         &+0.86(0.53)           &+3.56(0.002)           &+13.37(5.2e-6)       &+9.44(1.1e-5)            \\ 
\hline
Distilbert+adapted                                     &+0.45(0.04)           &+0.012(0.67)         &+0.11(0.68)           &+3.4(1.5e-7)           &+12.78(2.2e-8)       &+9.02(2.8e-6)           \\ 
\hline
\multicolumn{1}{l}{\textbf{\textbf{Severity}}}         & \multicolumn{1}{l}{}     & \multicolumn{1}{l}{}   & \multicolumn{1}{l}{}     & \multicolumn{1}{l}{}     & \multicolumn{1}{l}{}   & \multicolumn{1}{l}{}      \\ 
\hline
Bert+adapted                                           &+0.07(0.48)           &+5.96(0.0015)        &+1.16(0.0052)          &+0.14(0.42)             &+6.83(0.0015)       & 1.5(0.008)               \\ 
\hline
Distilbert+adapted                                     &+0.17(0.1)              &+1.44(0.0029)         &+1.04(0.08)             &+0.183(0.065)           &+1.45(0.0024)         &+0.74(0.007)           \\ 
\hline
\multicolumn{1}{l}{\textbf{\textbf{Type}}}             & \multicolumn{1}{l}{}     & \multicolumn{1}{l}{}   & \multicolumn{1}{l}{}     & \multicolumn{1}{l}{}     & \multicolumn{1}{l}{}   & \multicolumn{1}{l}{}      \\ 
\hline
Bert+adapted                                           &+2.82(0.0041)           &+4.72(0.0043)        &+3.14(0.0097)           &+49.48(2.8e-6)          &+68.91(3.5e-10)      & 63.44(2.8e-8)            \\ 
\hline
Distlbert+adapted                                      &+2.35( 0.0023)          &+3.69(0.0005)         &+2.52(0.0033)           &+50.57(2.5e-9)          &+70.88(3.5e-10)       &+64.92(2.8 e-8)          \\ 
\hline
% \cline{1-1}
\end{tabular}
}
\end{table}

The performance improvement table suggests that adapted models consistently have positive contributions. 
That is, the overall performance of the adapted models is promising. 
However, positive differences for certain cases are not found to be statistically significant as seen in the table since the corresponding $p$-values are not smaller than 0.05.

%%%%%%%%%%%%%%%%%%%%%%%%%%%%%%%%%%%%%%%%%%%%%%%%%%%%%%%%%%%%%%%%%%%%%
% \subsection{Analysis of adaptive fine-tuning}
%%%%%%%%%%%%%%%%%%%%%%%%%%%%%%%%%%%%%%%%%%%%%%%%%%%%%%%%%%%%%%%%%%%%%
We can clearly see that \textit{Priority} and \text{Type} tasks have a strong pattern between requirement and the target class. Therefore, we observe better improvement for these tasks. However, \textit{Severity} class is the most difficult one to be accurately classified among these tasks, % problems. 
mostly because we are unable to observe a strong relationship between the requirement text and the target. 
In many models and many different experimental setups that we have designed, no significant pattern have been observed for this class. Therefore, the Transformer models are not able to achieve any meaningful improvement against the baseline methods, nor do we see an improvement during adaptive fine-tuning. The feature extraction-based models (Word embeddings and SBERT) show better performance especially in terms of F1-score for this task.

%%%%%%%%%%%%%%%%%%%%%%%%%%%%%%%%%%%%%%%%%%%%%%%%%%%%%%%%%%%%%%%%%%%%%
\subsection{Detailed performance evaluation results} % with DOORS data}
%%%%%%%%%%%%%%%%%%%%%%%%%%%%%%%%%%%%%%%%%%%%%%%%%%%%%%%%%%%%%%%%%%%%%
% \tcolR{Using the best performing model (e.g., distillbert) ADD Detailed performance  values over individual classes -- see Table~4 in Derya's cascon paper  - DONE}

The confusion matrix and certain other metrics such as precision and recall can provide better insights to understand the prediction errors. 
We check this through the DistilBERT model, which is selected as the representative Transformer model. 
It is important to note that we observe similar behavior in other models as well, and DistilBERT is simply chosen as a representative model.
Table~\ref{tab:detailed_perf} shows detailed performance values for the adaptively fine-tuned version of DistilBERT. 
Based on these results, we identify one of the main reasons for the errors in Priority classification as the model being too sensitive to high-frequency labels such as ``Unassigned''. 
That is, the performance for low-frequency labels ``Medium'' and ``Low'' suffer from the imbalanced class distribution. 

\begin{table}[!ht]
\centering
\caption{Detailed performance results for the adaptively fine-tuned DistilBERT model for the Priority, Severity, and Type prediction tasks (diagonals bolded in confusion matrices).}
\label{tab:detailed_perf}
    \-\hspace{-2.2cm}\subfloat[Priority performance values by class \label{tab:priority_perf}]
    {
    \resizebox{0.395\linewidth}{!}{
    \begin{tabular}[c]{|l|rrrr|}
    \hline
    & \textbf{Precision} & \textbf{Recall} & \textbf{F1-score} & \textbf{Support} \\
    \hline
    \textbf{High}       & 0.582 & 0.507 & 0.542    & 682     \\
    \textbf{Low}        & 1.000 & 0.022 & 0.044    & 132     \\
    \textbf{Medium}     & 0.466 & 0.191 & 0.271    & 497     \\
    \textbf{Unassigned} & 0.785 & 0.926 & 0.849    & 2,837   \\
    \hline
    \end{tabular}
    }
    }
    \subfloat[Priority confusion matrix \label{tab:priority_cm}] 
    {
    \resizebox{0.395\linewidth}{!}{
    \begin{tabular}[c]{|ll|rrrr|}
    \hline
    & & \multicolumn{4}{c|}{Prediction} \\
    % \hline
    & & \textbf{High} & \textbf{Low} & \textbf{Medium} & \textbf{Unassigned} \\
    \hline
    \multirow{4}{*}{\STAB{\rotatebox[origin=c]{90}{{Ground Truth}}}} & \textbf{High} & \textbf{346} & 0 & 37 & 299 \\
    & \textbf{Low} & 13 & \textbf{3} & 12 & 104 \\
    & \textbf{Medium} & 86 & 0 & \textbf{95} & 316 \\
    & \textbf{Unassigned} & 149 & 0 & 60 & \textbf{2,628}\\
    \hline
    \end{tabular}
    }
    }\\
    \-\hspace{-0.79cm}\subfloat[Severity performance values by class \label{tab:severity_perf}]
    {
    \resizebox{0.395\linewidth}{!}{
    \begin{tabular}[c]{|l|rrrr|}
    \hline
     & \textbf{Precision} & \textbf{Recall} & \textbf{F1-score} & \textbf{Support} \\
    \hline
    \textbf{Blocker} & 0.800 & 0.150 & 0.250 & 26 \\ 
    \textbf{Critical} & 1.000 & 0.066 & 0.125 & 60 \\ 
    \textbf{Major} & 0.390 & 0.195 & 0.262 & 291 \\ 
    \textbf{Minor} & 1.000 & 0.134 & 0.237 & 67 \\ 
    \textbf{Normal} & 0.890 & 0.977 & 0.933 & 3,572 \\ 
    \textbf{Undecided} & 0.695 & 0.431 & 0.532 & 132 \\ 
    \hline
    \end{tabular}
    }
    }\hspace{0.01cm}
    \subfloat[Severity confusion matrix\label{tab:severity_cm}] 
    {
    \resizebox{0.515\linewidth}{!}{
    \begin{tabular}[c]{|ll|rrrrrr|}
    \hline
     & & \multicolumn{6}{c|}{Prediction} \\
     & & \textbf{Blocker} & \textbf{Critical} & \textbf{Major} & \textbf{Minor} & \textbf{Normal} & \textbf{Undecided} \\
     \hline
    \multirow{7}{*}{\STAB{\rotatebox[origin=c]{90}{{Ground Truth}}}} & \textbf{Blocker} & \textbf{4} & 0 & 7 & 0 & 14 & 1 \\
 & \textbf{Critical} & 0 & \textbf{4} & 16 & 0 & 39 & 1 \\
 & \textbf{Major} & 1 & 0 & \textbf{57} & 0 & 232 & 1 \\
 & \textbf{Minor} & 0 & 0 & 2 & \textbf{9} & 56 & 0 \\
 & \textbf{Normal} & 0 & 0 & 59 & 0 & \textbf{3,491} & 22 \\
 & \textbf{Undecided} & 0 & 0 & 2 & 0 & 73 & \textbf{57}\\
    \hline
    \end{tabular}
    }
    }\\
        \subfloat[Type performance values by class \label{tab:type_perf}]
    {
    \resizebox{0.395\linewidth}{!}{
    \begin{tabular}[c]{|l|rrrr|}
    \hline
     & \textbf{Precision} & \textbf{Recall} & \textbf{F1-score} & \textbf{Support} \\
    \hline
    \textbf{Enhancement} & 0.763 & 0.823 & 0.792 & 1,251 \\
    \textbf{JUnit} & 0.922 & 0.902 & 0.912 & 132 \\
    \textbf{Maintenance} & 0.825 & 0.895 & 0.858 & 562 \\
    \textbf{Other} & 0.589 & 0.354 & 0.442 & 466 \\
    \textbf{Plan Item} & 0.749 & 0.809 & 0.778 & 199 \\
    \textbf{Story} & 0.916 & 0.932 & 0.924 & 1,131 \\
    \textbf{Test Task} & 0.920 & 0.936 & 0.928 & 407 \\
    \hline
    \end{tabular}
    }
    }
    \subfloat[Type confusion matrix \label{tab:type_cm}] 
    {
    \resizebox{0.605\linewidth}{!}{
    \begin{tabular}[c]{|ll|rrrrrrr|}
    \hline
     & & \multicolumn{7}{c|}{Prediction} \\
     & & \textbf{Enhancement} & \textbf{JUnit} & \textbf{Maintenance} & \textbf{Other} & \textbf{Plan Item} & \textbf{Story} & \textbf{Test Task} \\
     \hline
    \multirow{7}{*}{\STAB{\rotatebox[origin=c]{90}{{Ground Truth}}}} & \textbf{Enhancement} & \textbf{1,029} & 1 & 70 & 80 & 21 & 48 & 2 \\
     & \textbf{JUnit} & 1 & \textbf{119} & 1 & 2 & 1 & 6 & 2 \\
     & \textbf{Maintenance} & 44 & 0 & \textbf{503} & 10 & 1 & 4 & 0 \\
     & \textbf{Other} & 212 & 3 & 33 & \textbf{165} & 12 & 17 & 24 \\
     & \textbf{Plan Item} & 18 & 0 & 0 & 7 & \textbf{161} & 12 & 1 \\
     & \textbf{Story} & 41 & 1 & 2 & 11 & 18 & \textbf{1,054} & 4 \\
     & \textbf{Test Task} & 4 & 5 & 1 & 5 & 1 & 10 & \textbf{381} \\
    \hline
    \end{tabular}
    }
    }
\end{table}

The severity classification task shows similar behavior to the Priority category. 
This can be similarly attributed to severe data imbalance for this label, with the ``Normal'' class having a significant majority over the others.
We find that the predictions for the ``Major'' class can be quite poor, with many ``Major'' severity requirements predicted as ``Normal'' severity. 
On the other hand, the Type label has relatively better class distributions than others.
We observe that fine-tuning performs fairly well on this task. 
There is only one class label, ``Other'',  that performs significantly poorly than the others.

%%%%%%%%%%%%%%%%%%%%%%%%%%%%%%%%%%%%%%%%%%%%%%%%%%%%%%%%%%%%%%%%%%%%%
\subsection{Discussion on model predictions}
%%%%%%%%%%%%%%%%%%%%%%%%%%%%%%%%%%%%%%%%%%%%%%%%%%%%%%%%%%%%%%%%%%%%%
We next examine the sample requirements where the model fails to predict the actual class correctly in our three classification tasks with the DOORS dataset (see Table~\ref{tab:sample_results}). % for the purpose of investigating the model predictions over the test set for each classification task. 
Below, we summarize the general observations regarding these misclassifications.
\begin{itemize}\setlength\itemsep{0.3em}
    \item In general, the model predictions are highly impacted by class distributions and input length. The majority of the test instances are classified into ``Unassigned'', ``Normal'' and ``Enhancement'' classes for priority, severity, and type classification tasks, respectively. 
    
    \item In the Type classification task, the ``Story'' class is usually assigned to the instance that mentions a requirement for a particular software whereas the ``Enhancement'' class is usually assigned to an already existing task which mentions an improvement over the existing requirement. 
    Certain words such as  ``improve'', ``ensure'', and ``report'' might overlap in multiple classes which may lead to misclassification.
    
    \item Severity labels generally point to the extent of a particular defect or a task can impact the software product.
    However, the level of severity designation can be subjective.
    As such, the requirement text with the ``Critical'' or ``Major'' label can be classified as ``Normal'' because the model is not able to differentiate whether that particular requirement belongs to the important feature and use-case in the system.
    
    \item Priority classification task can be considered as ordering the requirements/defects as ``High'', ``Low'', and ``Medium'' based on business needs and urgency of solving a particular defect. 
    Similar to the Severity label, these priority values can also be highly subjective.
    Furthermore, since the ``Unassigned'' class is the significant majority, and may contain instances from the other three classes, Priority classification can be considered a more challenging task than others.
    % Generally, High, Medium, and Low priority tasks always starts with a verb such as `Support',`Ensure' and `Perform', however these verbs are also mentioned in the `Unassigned' tasks which creates the possibility of misclassification.
\end{itemize}

% Table~\ref{tab:sample_results} shows the examples of misclassification in each classification task. 
Table~\ref{tab:sample_results} provides specific examples of misclassification cases.
In priority classification, $R_1$ is classified into ``Unassigned'' (indicating appropriate priority level could not be determined), however, the requirement text indicates that it is a high priority task as the rename commands are not able to refresh the expected feature. 
Similarly, $R_2$ is also classified as ``Unassigned'' despite having lower priority. 
In Severity classification, $R_3$ is predicted as ``Normal'' although it is evident from the requirement text that the label should be ``Critical'', as the user cannot access graphical editor plug-ins on Linux and Mac workstations which might cause a complete halt in the application for Mac and Linux users. %, so it should be classifed as `Critical'. 
$R_4$ also suggests an important task as the user is not able to lock the changeset deliveries at the appropriate time and this task needs urgent attention, which explains the `Major' severity level. 
However, the model predicts the ``Normal'' level of severity. 
In Type classification, the $R_5$ task talks about the improvement in performance for some baseline query which can be easily identified as the maintenance task in the existing system. 
Words like ``improve'' might have confused the model, and it classifies $R_5$ as ``Story''. 
The presence of the word ``Assessment'' in $R_6$ gives a clear indication that it is a ``Test Task'', whereas the word ``web client'' also shows the presence of the user. 
Accordingly, the model is unable to differentiate between a ``Story'' and ``Test Task'' class in this case. 

\begin{table}[!ht]
    \centering
    \caption{Examples of misclassified instances}
    \label{tab:sample_results}
    \resizebox{1.0\linewidth}{!}{
    \begin{tabular}{P{0.15\textwidth} P{0.57\textwidth} P{0.12\textwidth} P{0.15\textwidth}}
    \toprule
    \textbf{Classification task} & \textbf{Requirement text} & \textbf{Actual class} & \textbf{Predicted class} \\ 
    \midrule
    Priority & $R_1$: Type system rename commands do not refresh the affected shapes in the TRS feed.  & High & Unassigned \\ 
    \cmidrule(l){2-4} 
    & $R_2$: Review email can be directed to user who does not have access to the RRC project. & Low  & Unassigned \\ 
    \midrule
    Severity & $R_3$: Support browser graphical editor plug-ins on Linux and Mac workstations & Critical & Normal \\ 
    \cmidrule(l){2-4} 
    & $R_4$: Changeset deliveries do not lock at appropriate time & Major & Normal \\ 
    \midrule
    Type & $R_5$: Improve performance for large baseline compare query & Maintenance Item & Story \\
    \cmidrule(l){2-4} 
    & $R_6$: Assessment of existing web client TCs & Test Task & Story \\ 
    \bottomrule
    \end{tabular}
    }
\end{table}

%%%%%%%%%%%%%%%%%%%%%%%%%%%%%%%%%%%%%%
\section{Conclusion}\label{sec:conclusion}
%%%%%%%%%%%%%%%%%%%%%%%%%%%%%%%%%%%%%%
Modern language models may still be inadequate to deal with transferring knowledge where the distribution of the target domain may differ significantly from that of the source domain. 
Even though it is well known that the Transformers are more robust to the OOD problem than other deep learning architectures, there is still some room for improvement in this regard. 
In this paper, we show that adaptive fine-tuning can help various text classification tasks in the SRS domain. 
In a three-stage training pipeline, we fine-tune the pre-trained models on data that is closer to the distribution of the target data, which benefits to reduce the variation in the source and target distributions.
For this method, it is sufficient to have unlabeled data from the target distribution as the model is fine-tuned with the pre-training objective.
We leverage the domain-adaptive fine-tuning for three problems of SRS document classification (Priority, Severity and Type classification), and improve the model performance on a real distribution shift. 
We find that the Severity classification is the most challenging task as this category does not contain salient patterns between the requirement text and the task. 
We perform comparisons against strong baselines and show that the model performance can be improved significantly by the additional adaptive phase. 

Our work can be extended in multiple directions. First, due to a lack of publicly available data, we only experimented with a single data source. 
The effectiveness of adaptive fine-tuning for SRS classification tasks can be further investigated by using diverse datasets. Second, we note that all three classification tasks considered in this study suffer from data imbalance issues.
Accordingly, data augmentation strategies can be employed alongside adaptive fine-tuning for improved classification performance.
Lastly, baseline models are all pre-trained on a general corpus, whereas a software engineering corpus-specific Transformer model training can perform better for our classification tasks.

\section*{Statements and Declarations}
No potential conflict of interest was reported by the authors.

\section*{Data Availability Statement}
The DOORS dataset is propriety and is not made available public. % are not shared due to confidentiality reasons.

% %\bibliographystyle{ACM-Reference-Format}
% % \printbibliography
% \bibliography{WileyNJD-AMA}
%%%%%%%%%%%%%%%%%%%%%%%%%%%%%%%%%%%%
%\bibliographystyle{ACM-Reference-Format}
\bibliographystyle{elsarticle-num-names}
\bibliography{main_springer}
%%%%%%%%%%%%%%%%%%%%%%%%%%%%%%%%%%%%

\end{document}